\documentclass[aps,prl,twocolumn,showkeys,showpacs,superscriptaddress,groupedaddress]{revtex4}  
\pdfoutput=1
\usepackage{graphicx}  
\usepackage{dcolumn}   
\usepackage{bm}        
\usepackage{amssymb}   
\usepackage{setspace}
\usepackage{color}
\usepackage{amsmath}
\usepackage{epstopdf}
\begin{document}
\begin{spacing}{1.0}
\widetext
\leftline{Primary authors: L.M. Yu}
\leftline{Current version 4.0}
\rightline{Email: zqiu@zju.edu.cn, chenw@swip.ac.cn}


\title{Experimental Evidence of Nonlinear Avalanche Dynamics of Energetic Particle Modes}

%

\affiliation{Southwestern Institute of Physics, P.O.Box 432, Chengdu 610041, China.}
\affiliation{Center for Nonlinear Plasma Science and ENEA C. R. Frascati, Via E. Fermi 45, 00044 Frascati (Roma), Italy.}
\affiliation{Institute for Fusion Theory and Simulation, Zhejiang University, Hangzhou 310027, China.}
\affiliation{Key Laboratory of Optoelectronic Devices and Systems of Ministry of Education and Guangdong Province, College of Physics and Optoelectronic Engineering, Shenzhen University, Shenzhen 518060, China.}
\affiliation{School of Physics and Optoelectronic, Dalian University of Technology, Dalian 116024, China.}

\author{L.M. Yu$^1$, F. Zonca$^{2,3}$, Z.Y. Qiu$^{*3}$, L. Chen$^3$, W. Chen$^{*1}$, X.T. Ding$^1$, X.Q. Ji$^1$, T. Wang$^{4}$, T.B. Wang$^1$, R.R. Ma$^1$, B.S. Yuan$^1$, P.W. Shi$^5$, Y.G. Li$^1$, L. Liu$^1$, Z.B. Shi$^1$, J.Y. Cao$^1$, J.Q. Dong$^1$, Yi Liu$^1$, Q.W. Yang$^1$ and M. Xu$^1$}

\noaffiliation
\vskip 0.25cm
\date{\today}

\begin{abstract}
Recent observations in HL-2A tokamak give new experimental evidences of energetic particle mode (EPM) avalanche. In a strong EPM burst, the mode structure propagates radially outward within two hundred Alfv{\'e}n time, while the frequency of the dominant mode changes self-consistently to maximize wave-particle power exchange and mode growth.
This suggests that significant energetic particle transport occurs in this avalanche phase, in agreement with theoretical framework of EPM convective amplification. A simplified relay runner model yields satisfactory interpretations of the measurements.
The results can help understanding the nonlinear dynamics of energetic particle driven modes in future burning plasmas, such as ITER.
\end{abstract}


\maketitle


\textbf{\textit{Introduction}}--The nonlinear interactions between energetic particle (EP) and Alfv{\'e}n waves are very important for astrophysics \cite{space} and high temperature plasmas \cite{3,wu}, especially for magnetic confinement fusion applications, because they will significantly affect the redistribution and transport of EPs. When the EPs have sufficiently strong pressure gradient, they can excite a non-normal mode, named energetic particle mode (EPM) \cite{1,2,3,4, 7}, which emerges as discrete fluctuation out of the shear Alfv{\'e}n continuous spectrum at the EP characteristic frequency for maximized wave-EP power exchange above the threshold condition due to continuum damping. Therefore, EP losses in a fusion device can be a more serious concern in the presence of EPMs. The frequency of the EPMs is determined by EP transit ($\omega_{tr}$), bounce ($\omega_{b}$), precessional ($\omega_{p}$) frequencies or their combination $\omega=n\omega_{p}+l\omega_{b,tr}+ \Delta$ (where, $n$ is toroidal mode number, $l=\pm1,~\pm2~\cdots$, and $\Delta = (n q - m) \omega_{tr}$ for circulating particles interacting with a dominant fluctuation with poloidal mode number $m$ and safety factor $q$).

The early nonlinear theories \cite{5,6,7,8,10b} of EPM have addressed and described the role of the EPs radial redistributions on the self-consistent nonlinear time evolution of such a strongly driven non-perturbative fluctuation structure. If the EPs move outward, they can locally alter the EP gradient in phase space and destabilize another fluctuation out of the shear Alfv{\'e}n wave continuum that transports them further \cite{3,8,9}. Qualitatively, we may think about this process as successive excitation of fluctuations out of the shear Alfv{\'e}n wave continuum much like what different runners do in a relay race.
Similar to runners that pass on a baton to their companion at the end of each leg to maximize the speed at which the finish line is reached, different fluctuations of the shear Alfv{\'e}n continuous spectrum become dominant at subsequent stages of the nonlinear dynamics to maximize wave particle power extraction \cite{3,8,10b,9,10}.
The simplest model description for such nonlinear evolution is the relay runner model (RRM) \cite{6}.
EPMs have been found on many devices during neutral beam injection (NBI) \cite{11,12,13,14,chs} and ion cyclotron resonant heating \cite{tftr}, causing redistribution and losses of EPs.
Some experimental results indicate the strong EPM produce convective radial redistribution \cite{11,12,chs}, and have been analyzed and simulated by theory \cite{6,18b,briguglio,AB1,AB2,AB3}.
However, up to now, the theoretically expected nonlinear evolution of EPM and the corresponding radial redistribution of EPs have not been unambiguously observed in experiments.
In this letter we will give novel evidences of this nonlinear avalanche dynamics from experimental observations with the fast magnetic probes and the tomography of soft X-ray (SXR) arrays in the HL-2A tokamak. From these experiments, the time evolution of the different modes can be observed clearly. The $m/n=2/1$ mode is excited firstly in the plasma core, then the $m/n=3/1$ and $4/1$ modes appear successively as the fluctuation structure propagates from plasma core to edge. The dependence of mode frequency chirping rate on mode amplitude and plasma parameters is in agreement with the RRM prediction and experimentally demonstrates that the EPM avalanche occurs non-adiabatically \cite{3,8,10b,9,10} as consequence of the non-perturbative interplay with EPs. The radial velocity of the EP convective transport is estimated from theory, which also agrees with the experimentally observed radial propagating velocity of the wave.

\textbf{\textit{Experimental Results}}--HL-2A is a mid size circular cross section tokamak with $1.65~m$ major and $0.4~m$ minor radius \cite{15}. The NBI co-injects into plasma with an injection angle of $58.1^{\circ}$, and energy of the deuterium beam ions ($E_b$) $\sim40-45~keV$. The power of NBI ($P_{NBI}$) can reach up to $1.5~MW$. Aboundant Alfv{\'e}nic waves have been observed during the NBI heating discharges \cite{16,17,18,19,20}.

\begin{figure}[!htbp]
\centering
\includegraphics[scale=0.25]{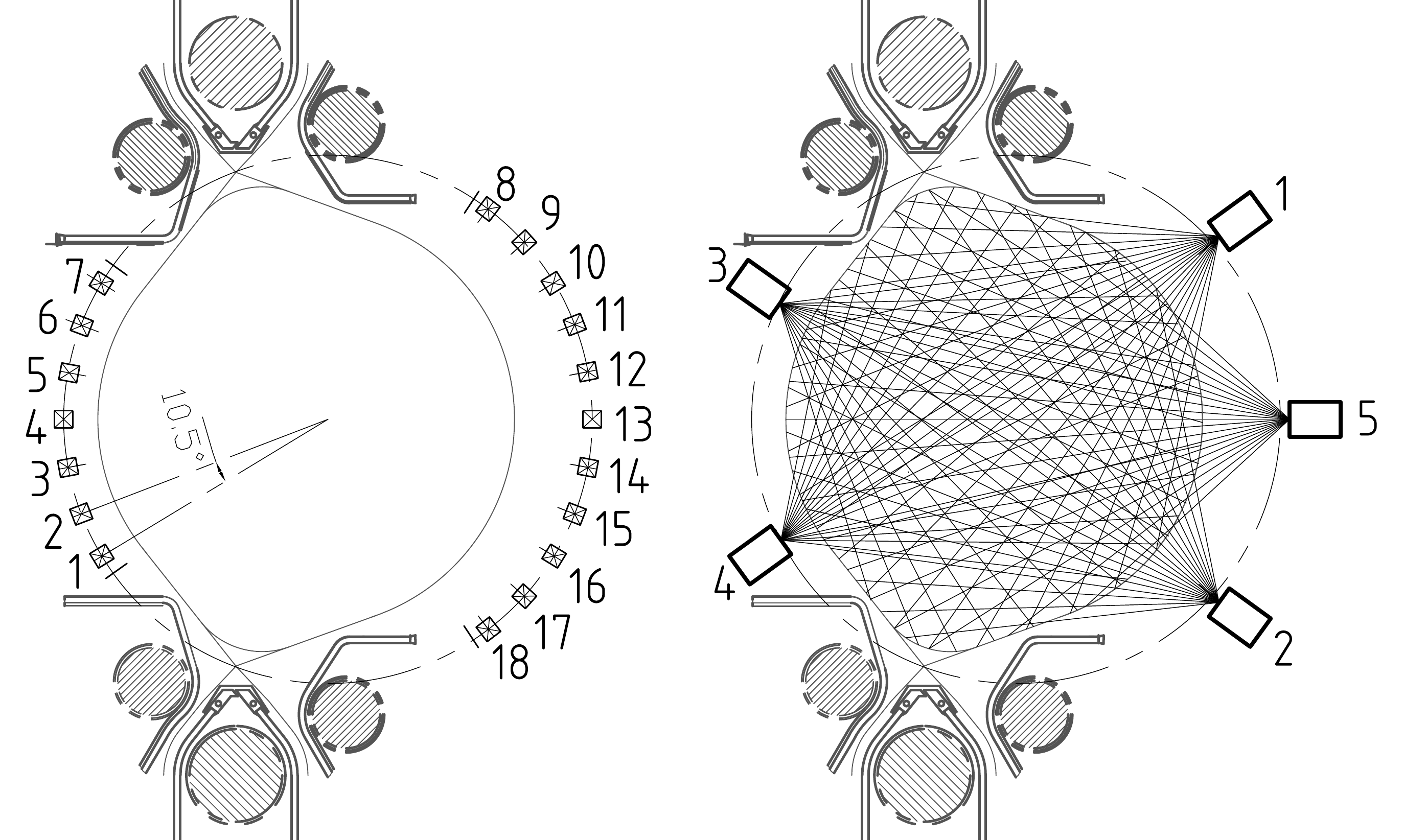}
\caption{\label{fig1} Arrangements of the poloidal Mirnov probes (left) and SXR arrays (right) in the HL-2A tokamak.}
\end{figure}

The Alfv{\'e}nic instabilities can be detected from the spectrogram of fast Mirnov probe and SXR array systems.
The poloidal and toroidal mode numbers of the wave are measured using the sets of $18$ poloidal (as shown in Fig.\ref{fig1} (left)) and $10$ toroidal Mirnov probe arrays, respectively.
There are five sets of SXR detection systems, each set consisting of $20$ evenly distributed arrays. The location and distribution of SXR arrays are shown in Fig.\ref{fig1} (right). 
The structure and evolution of the Alfv{\'e}nic modes inside $r=0.33~m$ region can be obtained by tomography \cite{24} of SXR arrays.

\begin{figure}[!htbp]
\centering
\includegraphics[scale=0.6]{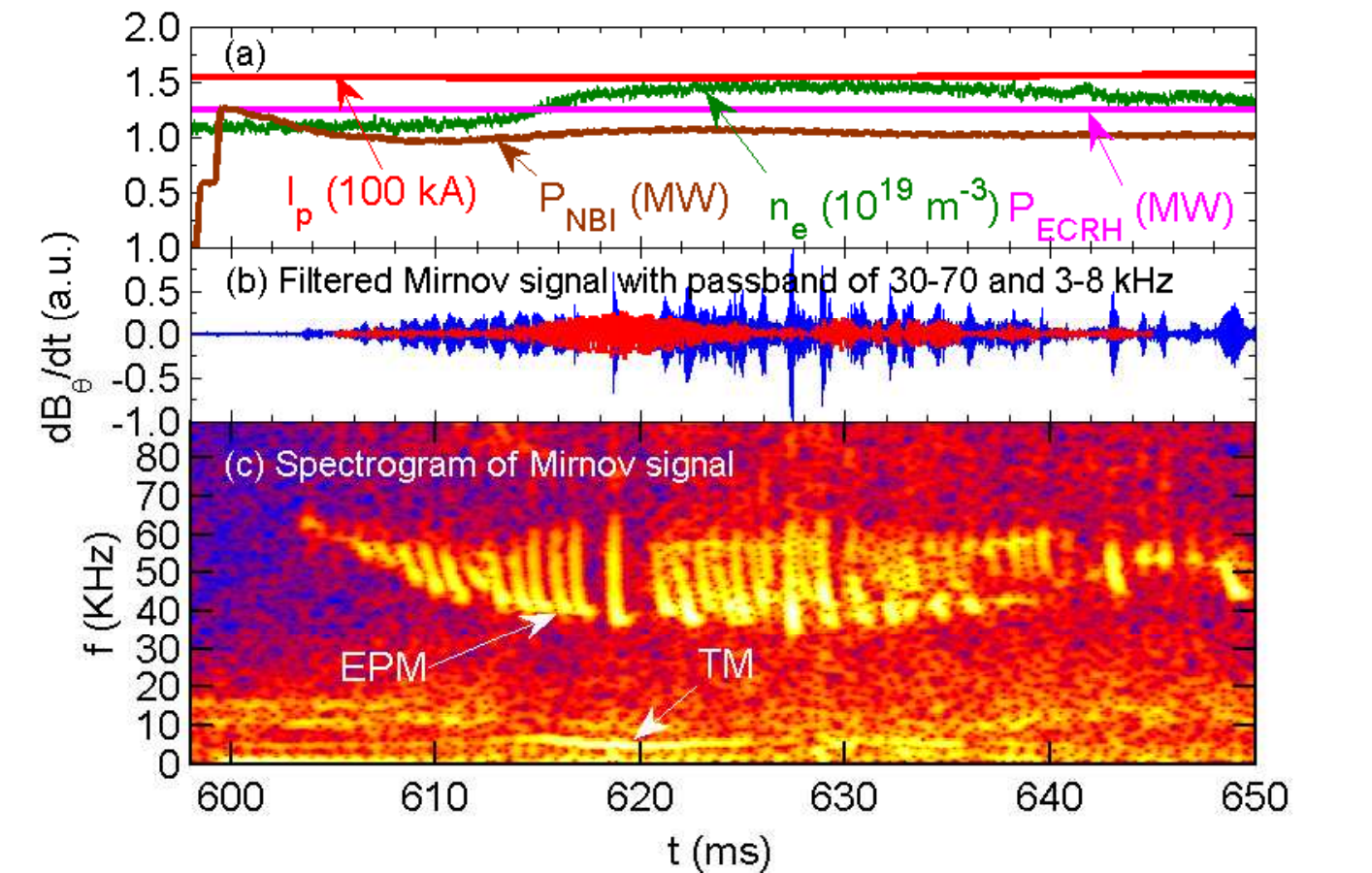}
\caption{\label{fig2}  EPMs observed in NBI heating plasma on HL-2A in shot I. (a) Discharge parameters,
(b) filtered Mirnov signal with passband of $30-75~kHz$ (blue waveform) and $3-8~kHz$ (red waveform), and (c) spectrogram of Mirnov singal.}
\end{figure}

There is a typical EPM event with frequency $f\sim35-65~kHz$ in shot I as shown in Fig.\ref{fig2}.
The discharge parameters are plasma current $I_{p}\sim155~ kA$, line averaged electron density $n_{e}\sim(1.1-1.5)\times10^{19}~m^{-3}$, toroidal magentic field $B_t=1.38~T$ and the $1.0~MW$ NBI with $E_{b}\sim42~keV$ switched on at $t=600~ms$, as shown in Fig.\ref{fig2} (a). Evident magnetic fluctuations of EPM are found on the poloidal Mirnov signal (No.12) after the injection of NBI, as shown by the blue waveform filtered with passband of $30-75~kHz$ in Fig.\ref{fig2} (b). The spectrogram of the Mirnov signal is shown in Fig.\ref{fig2} (c). The EPM frequency can sweep down from $65$ to $35~kHz$ within $\Delta t\sim1~ms$. There is a strong $f\sim3-8~kHz$ tearing  mode (TM) coexisting with EPM during $t=613-635~ms$. The fluctuations caused by TM are presented by the red waveform in Fig.\ref{fig2} (b).

\begin{figure}[!htbp]
\centering
\includegraphics[scale=0.55]{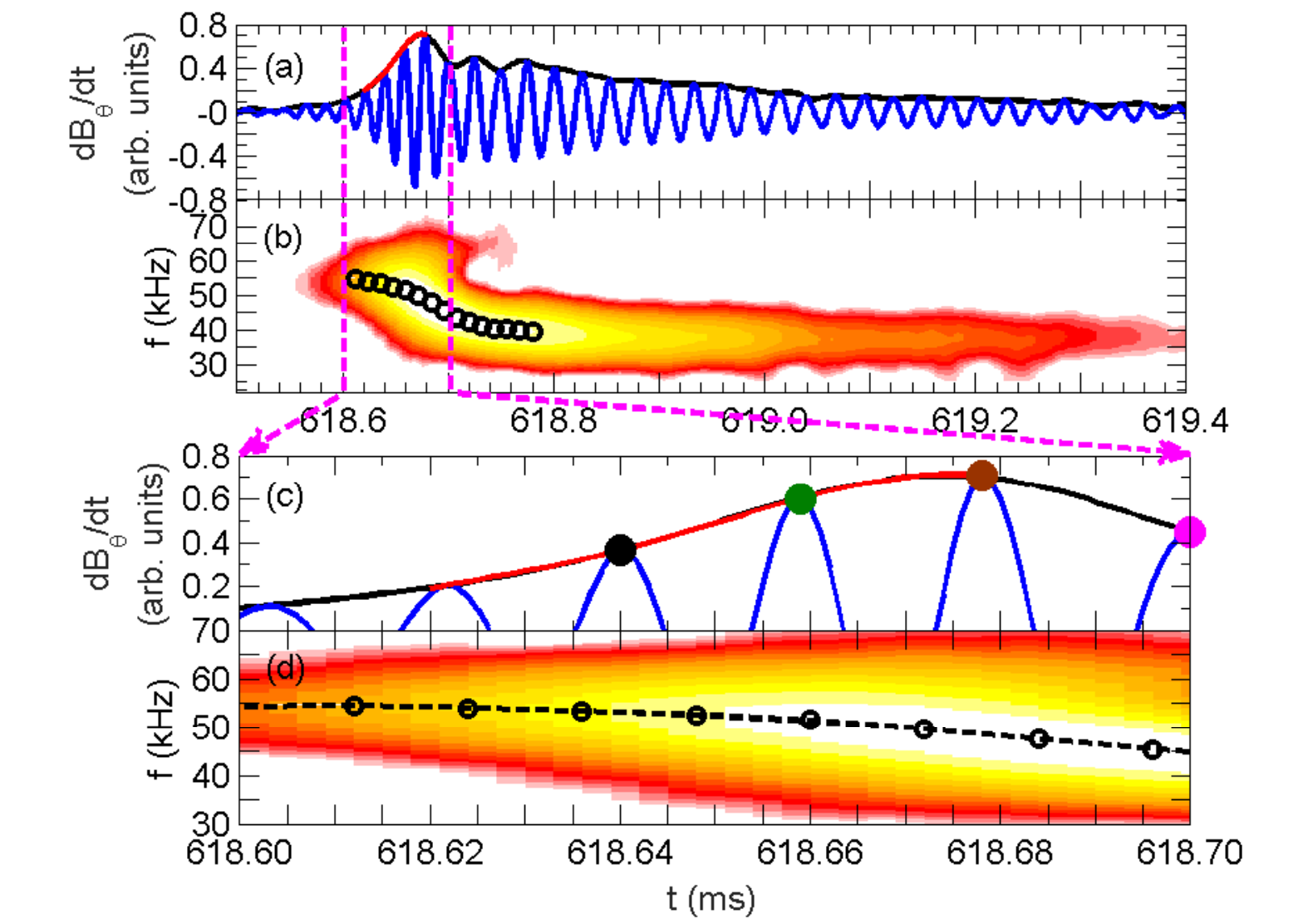}
\caption{\label{fig3} (a) Poloidal Mirnov signal (No.12) and its envelope line (red and black lines), (b) spectrogram of Mirnov signal and the mode frequency (black circles), the detail evolutions of (c) polodial Mirnov signal and its envelope (red line), and (d) mode frequency (black dashed circle line).}
\end{figure}

The filtered poloidal Mirnov signal of a strong EPM burst and its envelope labeled by the black line at $t=618.50-619.40~ms$ in shot I are shown in Fig.\ref{fig3} (a). The spectrogram of the Mirnov signal is shown in Fig.\ref{fig3} (b).
It is obvious that the amplitude of Mirnov signal, represented by the envelope, increases rapidly to the maximum during $t=618.62-618.68~ms$. The rising amplitude is identified by the red envelope line, as shown Fig.\ref{fig3} (c). The mode frequency drops from $55$ to $47~kHz$ during this phase, as shown in Fig.\ref{fig3} (d). Then, the amplitude decreases slowly, and the frequency continuously decreases from $47$ to $37~kHz$ during $t=618.68-619.40~ms$. The maximum fluctuation of $\delta B/B$ can reach to $\sim4.5\times10^{-4}$ near $t=618.68~ms$, indicating significant EP transport \cite{5,12,21}. The typical Mirnov singal growth time is $\sim200$ Alfv{\'e}n times, expecting convective EP redistributions can take place \cite{2,7,8,9,22}.

\begin{figure}[!htbp]
\centering
\includegraphics[scale=0.6]{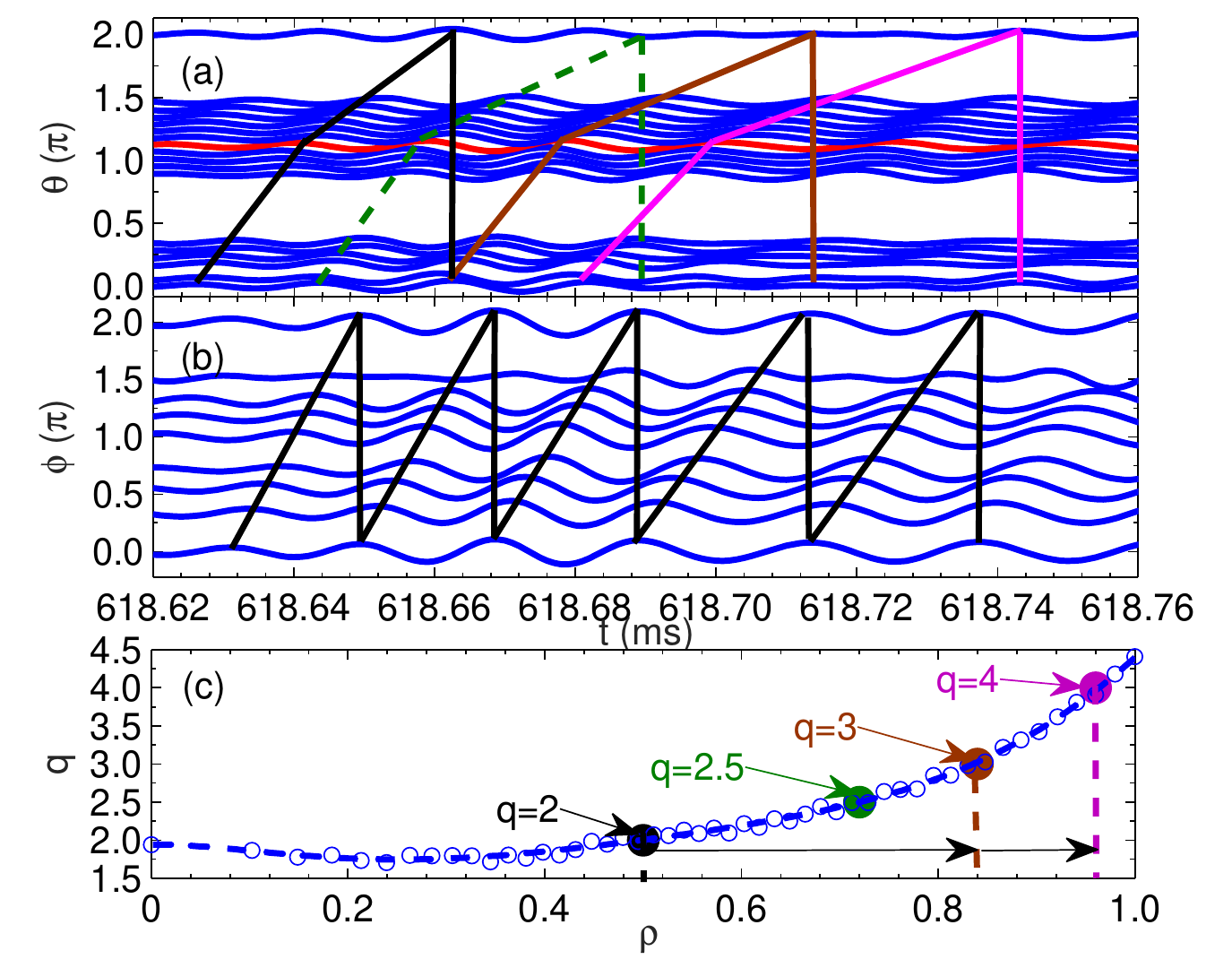}
\caption{\label{fig4} Evolution of the mode numbers of EPM. (a) Poloidal mode number changing from $m=2$ to $4$ successively, while (b) toroidal mode number keeping constant as $n=1$, (c) q-profile near $t=618~ms$ and the rational surface of $q=2$, $2.5$, $3$ and $4$ labeled by the black, green, brown and pink circles, respectively.}
\end{figure}

The evolution of mode numbers is confirmed by the phase shift of the waves from Mirnov probe arrays. There are the filtered polodial and toroidal Mirnov singals arranged from the bottom to the top at $t=618.62-618.76~ms$ in Fig.\ref{fig4} (a) and (b), respectively. The red wave line in Fig.\ref{fig4} (a) is the analyzed poloidal Mirnov signal, which is also shown in Fig.\ref{fig2} (b), Fig.\ref{fig3} (a) and (c). The mode propagates in ion diamagnetic drift direction poloidally. The poloidal mode number changes rapidly from $m=2$ to $3$, and then becomes $4$ successively, which are labeled by the black, brown and pink lines in Fig.\ref{fig4} (a), respectively. The green dashed line shows the process during which the dominant mode number changes from $m=2$ to $3$. At last, the m keeps as $4$ till the end the EPM event. The change of $m$ from $2$ to $4$ takes place within $\Delta t=0.06~ms$ (between $t=618.64-618.70~ms$), and the corresponding frequency changes from $53$ to $45~kHz$ during that time interval. The toroidal mode number is confirmed as $n=1$ during the whole chirping down process, as shown in Fig.\ref{fig4} (b). It indicates that the EPMs move from the core ($q=m/n=2$) to the edge ($q=3$ and $4$) of plasma.
The mode numbers and propagation direction of TM are confirmed as $m/n=3/1$ and in electron diamagnetic drift direction.
The q-profile is obtained by the current profile fitting method \cite{23}, as shown in Fig.\ref{fig4} (c).
The radial positions of the $q=2$, $2.5$, $3$ and $4$ rational surfaces are labeled by the black, green, brown and pink circles, respectively.

\begin{figure}[!htbp]
\centering
\includegraphics[scale=0.55]{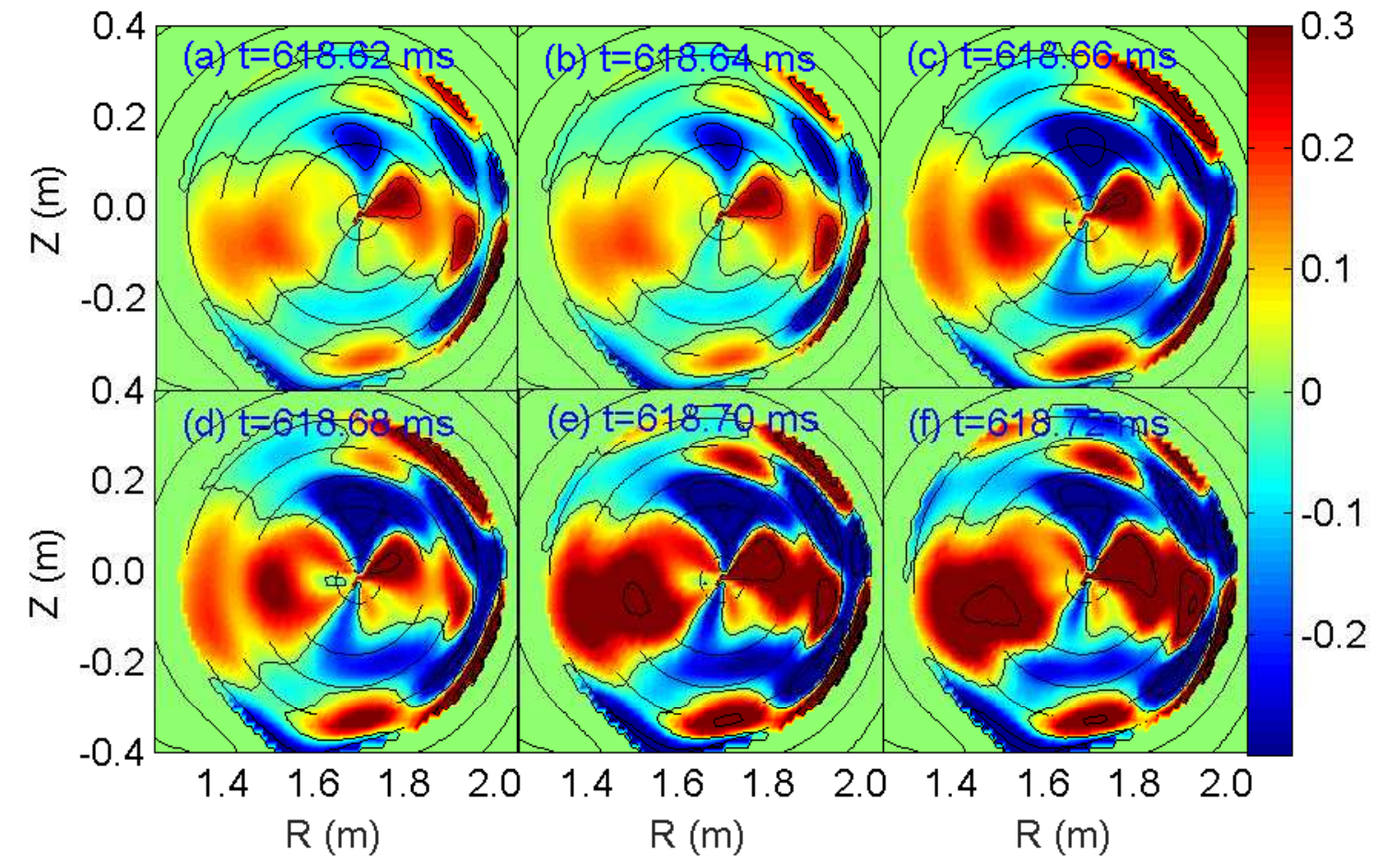}
\caption{\label{fig5} Mode structures and evolutions of EPM in poloidal cross section inside $r=0.33~m$ region obtained by tomography of SXR arrays.}
\end{figure}

The mode structures and evolutions of EPM and TM in poloidal cross section inside $r=0.33~m$ region at $t=618.62-618.72~ms$ are obtained by tomography of SXR arrays, as shown in Fig.\ref{fig5} (a)-(f). The $m=2$ mode mainly locates in the core of plasma at first, and then the mode propagates to the edge gradually. At last, the edge (parts of $m=3$ and $4$) elements become dominant.
The $m=3$ TM locates in the edge all the time.
The radial propagating processes of the EPM found from tomography are consistent with the changing tendency of mode numbers obtained by the Mirnov probe arrays.

\textbf{\textit{Theory Analysis and Verification}}--The nonlinear EPM evolution, including frequency chirping and radial propagation, is the result of EP radial transport and cooling by EPM. The EPM frequency is nonperturbatively determined by the EP characteristic frequency, while resonant EPs that lose energy during excitation of EPM will move outward and remain in resonance with the frequency downward chirping EPM, leading to convective EP transport and self-consistent EPM evolution.
Owning to the injection angle of NBI system on HL-2A, the EPM is driven by the transit resonance of the predominantly circulating EP population, i.e., $\omega=\omega_{tr}\equiv \sqrt{2(E/m_i-\mu B)}/(qR_0)$, with $E$, $m_i$ being respectively, the energy and mass of deuterium ions, and $\mu\equiv v^2_{\perp}/(2B)$ is the magnetic moment.
The self-consistent non-adiabatic EPM  frequency chirping, is determined by the phase-locking between the mode and resonant EPs that maximizes the power exchange, i.e.,
\begin{eqnarray}
\dot{\omega}=\dot{\omega}_{tr}\simeq \frac{\partial\omega_{tr}}{\partial E}\dot{E}+\frac{\partial\omega_{tr}}{\partial r} \dot{r},  \label{eq:dotwtr}
\end{eqnarray}
with the two terms on the right hand side representing EP transit frequency changing rate due to EP energy change and radial transport, respectively.

For a single-n coherent fluctuation of interest here, one obtains $\dot {E}\simeq (\omega/n)\dot{P}_{\phi}$ \cite{8,22} from the conservation of extended phase space Hamiltonian, with $P_{\phi}$ being the toroidal angular momentum and $\dot{P}_{\phi}\simeq -m_i\Omega_i r\dot{r}/q$. For the HL-2A parameters, the chirping rate contribution due to radial transport is much smaller than that due to energy change by $O(\omega_{*EP}/\omega)\sim 3.5  \times 10^{-2}$, with $\omega_{*EP}$ the EP diamagnetic frequency, indicating that the observed EPM on HL-2A are predominantly excited via velocity space anisotropy. One then has the approximate frequency chirping rate
\begin{eqnarray}
\dot{\omega}=\dot{\omega}_{tr}\simeq -\frac{\Omega_i r}{nR^2_0q^3}\dot{r}.  \label{eq:dotw}
\end{eqnarray}

On the other hand, the radial  velocity of resonant EPs can be derived as
\begin{eqnarray}
\dot{r}=\frac{c}{B_0}\mathbf{\hat{r}\cdot}\hat{\mathbf{b}}\times\mathbf{\delta E_{\theta}}\simeq \frac{k_{\theta}}{B_0k_{\parallel}k_r} A. \label{eq:dotr}
\end{eqnarray}
Here,  $k$ is the WKB wavenumber,  $\delta E_{\theta}$ is the poloidal electric field perturbation, and $A$ is the time derivative of the perturbed poloidal magnetic field from Mirnov signal, i.e., $A=\partial_t \delta B_{\theta}$.
Substituting into equation (\ref{eq:dotw}), noting that $k_{\theta}\equiv nq/r$ and $k_{\parallel}\equiv (nq-m)/(qR_0)$, one obtains
\begin{eqnarray}
\dot{\omega}\simeq -\frac{\Omega_i}{q^2R^2_0B_0k_{\parallel} k_r} A\propto \frac{A}{q^{1+\alpha}}, \label{eq:dotomega}
\end{eqnarray}
with $|\alpha|\leq 1$ and $q^{-\alpha}$ accounting for the radial dependence of $nq-m$ and $k_r$. Note that, given $\Theta$, the wave-particle phase, the expression of $\dot \omega_{tr}$ in equation (\ref{eq:dotwtr}) also gives $\ddot \Theta$ for resonant particles. That is \cite{3,10}
\begin{equation}
\ddot \Theta \sim \frac{\partial \omega_{tr}}{\partial E} \dot E \sim \omega_B^2,  \label{eq:4*}
\end{equation}
with $\omega_B$ the wave-particle trapping frequency.
This implies that, if equation (\ref{eq:dotomega}) is verified, the wave-EP interaction in EPM nonlinear dynamics is a non-adiabatic process with $\dot \omega \sim \omega_B^2$, with important implications that are further discussed below.

\begin{figure}[!htbp]
\centering
\includegraphics[scale=0.6]{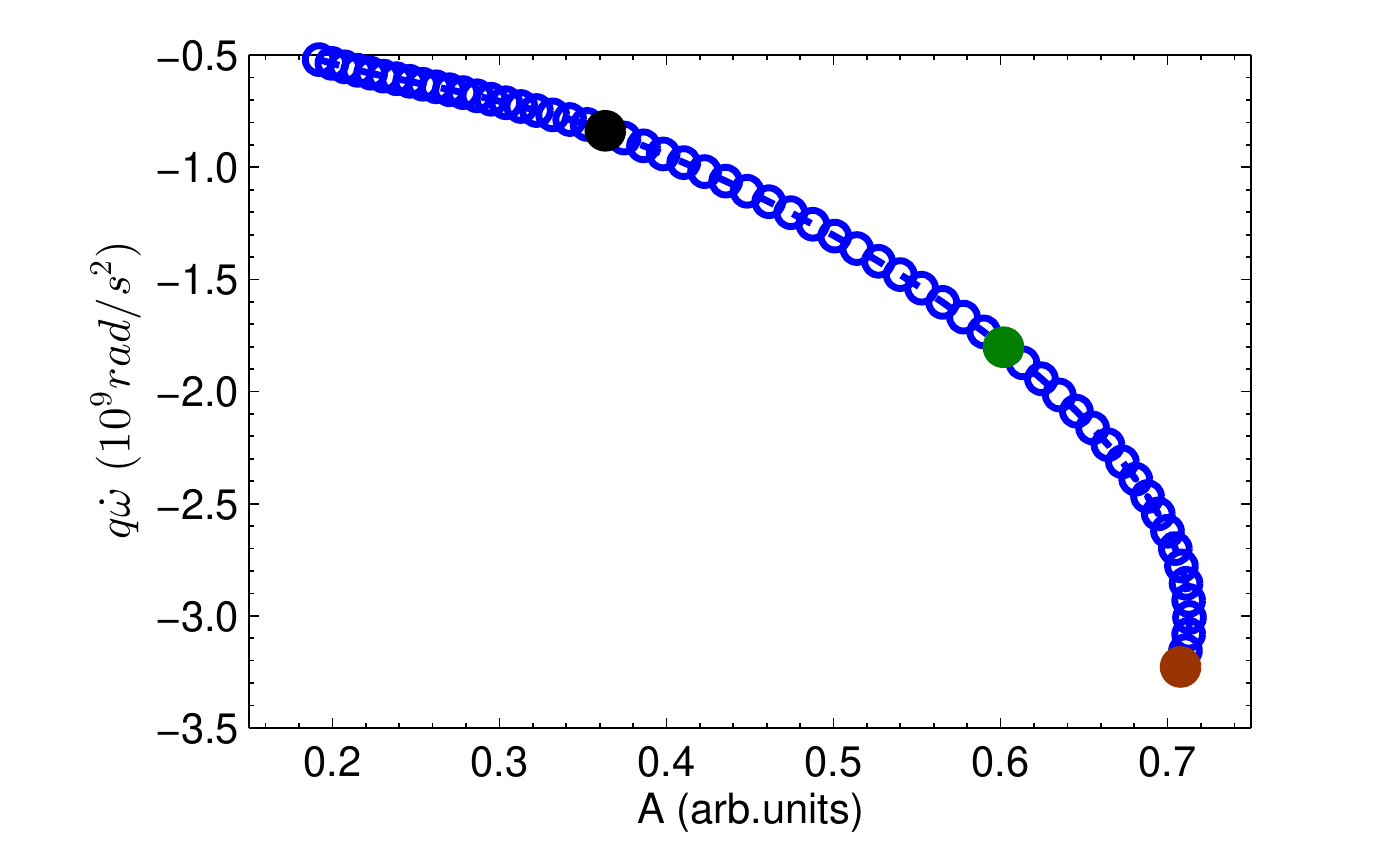}
\caption{\label{fig6} Relationship between $A$ and $q\dot{\omega}$ in experiment during $t=618.62-618.68~ms$. The black, green and brown circles present the relationship in $q=2$, $2.5$ and $3$ conditions.}
\end{figure}

The temporal evolution of the Mirnov signal ($A$) and frequency of EPM is shown in Fig.\ref{fig3} (c) and (d), respectively.
The $A$ (proportional to the mode amplitude), labeled by the red line in Fig.\ref{fig3}, increases during $t=618.62-618.68~ms$ when the mode  propagates from $q=2$ to $3$ surfaces ($\sim 1/3$ of HL-2A minor radius), showing convective amplification behavior typical of EPM nonlinear dynamics. $A$ then decreases during $t=618.68-618.70~ms$  as the mode propagates radially further from $q=3$ to $4$ surfaces, indicating  potentially serious EP losses as EPs reach plasma edge. Besides that, the $A$ decrease may also relate to the weakened drive due to phase-unlocking between the mode and EPs, and/or enhanced dissipation at the plasma edge. We will not focus on its mechanism here.

From equation (\ref{eq:dotomega}), the frequency chirping rate $\dot{\omega}$, $A$ and $q$ are connected by the scaling $A\propto q\dot{\omega}$, assuming $\alpha\ll1$ for simplicity. It can be seen that $A$ is roughly proportional to $q \dot{\omega}$, as shown in Fig.\ref{fig6}. The black, green and brown circles in Fig.\ref{fig6}, correspond  to the $q=2$, $2.5$ and $3$ conditions as shown in Fig.\ref{fig3} (c) and Fig.\ref{fig4} (c). The $A$ is perfectly proportional to $q \dot{\omega}$ when $A$ is smaller than $0.61$ (EPs propagate from $q=2$ to $2.5$ surfaces), showing strict consistency with the RRM prediction. As EPM propagates further, there is small deviation from the scaling $A\propto q\dot{\omega}$, as $A$ is in the range of $0.61-0.71$ (between the green and brown circles), caused by the mechanisms discussed above.

\begin{table}[!hbp]
\small
\centering
\caption{Experimental data and calculated results.}
\begin{tabular}{ccccccc}
\hline
\multicolumn{1}{c}{$q$}&\multicolumn{1}{c}{$t~(ms)$}&\multicolumn{1}{c}{$A~(a.u.)$}&\multicolumn{1}{c}{$r~(m)$}&\multicolumn{1}{c}{$f~(kHz)$}&\multicolumn{1}{c}{$\dot{\omega}~(10^9~rad/s^2)$}&\multicolumn{1}{c}{$\dot{r}~(km/s)$}\\
\hline
\multicolumn{1}{c}{$2$}&\multicolumn{1}{c}{$618.64$}&\multicolumn{1}{c}{$0.36$}&\multicolumn{1}{c}{$0.19$}&\multicolumn{1}{c}{$53.10$}&\multicolumn{1}{c}{$-0.42$}&\multicolumn{1}{c}{$0.78$}\\
\hline
\multicolumn{1}{c}{$2.5$}&\multicolumn{1}{c}{$618.66$}&\multicolumn{1}{c}{$0.61$}&\multicolumn{1}{c}{$0.27$}&\multicolumn{1}{c}{$51.27$}&\multicolumn{1}{c}{$-0.76$}&\multicolumn{1}{c}{$1.93$}\\
\hline
\multicolumn{1}{c}{$3$}&\multicolumn{1}{c}{$618.68$}&\multicolumn{1}{c}{$0.71$}&\multicolumn{1}{c}{$0.31$}&\multicolumn{1}{c}{$48.62$}&\multicolumn{1}{c}{$-1.06$}&\multicolumn{1}{c}{$4.00$}\\
\hline
\end{tabular}
\end{table}

The instantaneous radial velocity of the outward pumped EPs ($\dot{r}$) can be derived from equation (\ref{eq:dotw}) as $\dot{r}=-nR_{0}^2q^3\dot{\omega}/(\Omega_{i}r)$.
Substituting the experimental data into the equation, the obtained radial velocity at $q=2$, $2.5$ and $3$ rational surfaces and the corresponding experimental data are shown in Table.1. The averaged radial velocity of EPs ($\bar{\dot{r}}$) is about $2.23~km/s$.

On the other hand, the radial velocity of EPM ($V_p$) can be estimated by the ratio of the distance from $q=2$ to $3$ rational surfaces ($\Delta r$) to the time interval ($\Delta t$) as $V_p\simeq3.00~km/s$ according to the experimental data in Table.1, and one obtains $V_p\simeq \bar{\dot{r}}$.
This verifies experimentally equations (\ref{eq:dotwtr}) and (\ref{eq:dotw}), and, on the basis of equation (\ref{eq:4*}), it demonstrates that the EPM nonlinear dynamics is a non-adiabatic process \cite{3,10}. The crucial implication of this is that EP transport is not in buckets trapped by the wave \cite{Min94} but rather a convective process due to continuous trapping and de-trapping of particles that amplify the wave packet as it propagates radially. Analogies of this process with autoresonance in nonlinear dynamics and with superradiance in free-electron lasers are discussed in \cite{3,10}.
Therefore, the experimental evidence on HL-2A demonstrates for the first time that
the average radial velocity of the outward convected EPs agrees with the radial velocity of EPM from experiment, indicating the nonlinear mode evolution is the self-consistent result of EP radial transport by the EPM itself, as predicted by RRM model.

\textbf{\textit{Summary}}--The experimental evidence of nonlinear avalanche dynamics of the EPM is found in NBI heated plasma on HL-2A. The poloidal mode number of EPM changes from $m=2$ to $4$ successively, while the toroidal mode number always keeps as $n=1$, indicating that the EPM propagates from the core to edge gradually, covering a significant fraction of the tokamak minor radius.
The tomography of SXR arrays also illuminates the fast nonlinear avalanche. The product of $\dot{\omega}$ and $q$ factor is proportional to $A$ in the nonlinear avalanche process, e.g., $A\propto q\dot{\omega}$, which is consistent with the nonlinear RRM model prediction. The estimated radial velocity of the resonant EPs is close to the radial propagating velocity of the EPM in experiment, confirming  the convective amplification mechanism underlying EPM self-consistent nonlinear dynamics. Experimental evidence also supports the non-adiabatic nature of the EPM nonlinear dynamics connected with continuous trapping and de-trapping of EPs and their convective transport, similar to superradiance in free-electron lasers. This is the first time that experimental evidence of EP nonlinear avalanche dynamics, is demonstrated, consistent with theoretical predictions, which is of crucial importance for the understanding of transport and redistribution of EPs in future fusion reactor, such as ITER.

Two of the authors (L.M. Yu and W. Chen) are very grateful to the HL-2A Group. This work is supported in part by the National Key R$\&$D Program of China under Grant Nos. 2019YFE03020000, 2017YFE0301900, 2017YFE0301202 and 2018YFE0304102, by NNSF of China under Grant Nos. 11875024, 11875021 and 11835010, and by Sichuan Science and Technology Program Nos. 2020YFSY0047 and 2020JDJQ0070.

\end{spacing}

\end{document}